\documentstyle[12pt]{article}
\newlength{\extraspace}
\newlength{\extraspaces}
\newcommand{\be}{\begin{equation}
\addtolength{\abovedisplayskip}{\extraspaces}
\addtolength{\belowdisplayskip}{\extraspaces}
\addtolength{\abovedisplayshortskip}{\extraspace}
\addtolength{\belowdisplayshortskip}{\extraspace}}
\newcommand{\ee}{\end{equation}}

\newcommand{\ba}{\begin{eqnarray}
\addtolength{\abovedisplayskip}{\extraspaces}
\addtolength{\belowdisplayskip}{\extraspaces}
\addtolength{\abovedisplayshortskip}{\extraspace}
\addtolength{\belowdisplayshortskip}{\extraspace}}
\newcommand{\ea}{\end{eqnarray}}

\newcommand{\nonu}{\nonumber \\[.5mm]}
\newcommand{\A}{&\!\!\!}

\newcommand{\newsection}[1]{
\vspace{7mm} \pagebreak[3] \addtocounter{section}{1}
\setcounter{subsection}{0} \setcounter{footnote}{0}
\begin{center}
{\large {\bf \thesection. #1}}
\end{center}
\nopagebreak
\medskip
\nopagebreak \hspace{3mm}}

\begin{document}

\begin{center}
{{\bf Axially symmetric solution in Teleparallel Theory of
Gravitation}}
\end{center}
\centerline{ Gamal G.L. Nashed}

\bigskip

\centerline{{\it Mathematics Department, Faculty of Science, Ain
Shams University, Cairo, Egypt }}

\bigskip
 \centerline{ e-mail:nasshed@asunet.shams.eun.eg}

\hspace{2cm}
\\
\\
\\
\\
\\
\\
\\
\\

An exact solution  has an axial symmetry is obtained in the
teleparallel theory of gravitation.  The associated metric has the
structure function $G(\xi)=1-{\xi}^2-2mA{\xi}^3$. The cubic nature
of the structure function can make calculations cumbersome.  Using
a coordinate transformation we get a tetrad that its associated
metric has the structure function  in a factorisable form. This
new form has the advantage that its roots are now trivial to write
down. The singularities of the obtained tetrad are studied. Using
another coordinate transformation we get a tetrad that its
associated metric gives the Schwarzschild spacetime. Calculate the
energy content of this tetrad we get a meaningless result!
\newpage
\begin{center}
\newsection{\bf Introduction}
\end{center}

The C-metric is well known to describe a pair of black holes
undergoing uniform acceleration. It is usually written in a form
first given by Kinnersley and Walker \cite{KW} \be
ds^2=\displaystyle{ 1 \over A^2(x-y)^2} \left[ G(y)
dt^2-\displaystyle{dy^2 \over G(y)} + \displaystyle{dx^2 \over
G(x)}+G(x) d\phi^2 \right], \ee where the stricture function $G$
is defined by \be G(\xi)=1-\xi^2-2mA\xi^3.\ee Here $m$ and $A$ are
positive parameters related to the mass and acceleration of the
black hole, such that $mA<1/\sqrt{27}$. The fact that $G$ is a
cubic polynomial in $\xi$ means that one can  not in general write
down simple expression for its roots. Since these roots play an
important role in almost every analysis of the C-metric, most
results have to be expressed implicitly in   terms of them. Any
calculation which requires their explicit forms would naturally be
very tedious if not impossible to carry out \cite{FZ,B3,B4}.

Einstein's general relativity (GR) is very successful in
describing long distance (macroscopic) phenomena. This theory,
however, encounters serious difficulties on microscopic distances.
So far essential problems appear in all attempts to quantize the
standard GR \cite{Sc,Yi}. Also, the Lagrangian structure  of GR
differs, in principle, from the ordinary microscopic  gauge
theories. In particular, a covariant conserved energy-momentum
tensor for the gravitational field can not be constructed in the
framework of GR. consequently, the study of alternative models of
gravity is justified from the physical as well as from the
mathematical point of view. Even in the case when GR is unique
true theory of gravity, consideration of close alternative models
can shed light on the properties of GR itself.

Theories of gravity based on the geometry of distance parallelism
\cite{YJ}$\sim$\cite{Jm} are commonly considered as the closest
alternative to the general relativity theory. Teleparallel gravity
models posses a number of attractive features both from the
geometrical and physical viewpoints. Teleparallelism is naturally
formulated by gauging external (spacetime) translation and
underlain the Weitzenb$\ddot{o}$ck spacetime characterized by the
metricity condition and by the vanishing of the curvature tensor.
Translations are closely related to the group of general
coordinate transformations which underlies general relativity.
Therefore, the energy-momentum tensor represents the matter source
in the field equations of tetradic theories of gravity like in
general relativity.

It is the aim of the present work to derive an axially symmetric
solution in teleparallel theory of gravitation. In section 2 we
give brief review of the teleparallel theory of gravitation. A
tetrad having four unknown functions is applied to the field
equation of the teleparallel theory of gravity then, a solution of
axial symmetry is obtained in section 3. A coordinate
transformation is
 applied to the obtained tetrad in section 3, to put the structure function in a
factorisable form.  The advantage of this transformation is that
it makes the roots of the original solution be factorisable. In
section 4, the singularities of this tetrad are studied. In
section 5, another coordinate transformation is applied and a
tetrad that its associated  metric gives the Schwarzschild
spacetime is obtained. The energy content of this tetrad is also
calculated in section 5. Discussion and conclusion of the obtained
results are given in the final section.
\newsection{The  tetrad theory of gravitation}

In a spacetime with absolute parallelism the parallel vector
fields ${e_i}^\mu$ define the nonsymmetric connection \be
{\Gamma^\lambda}_{\mu \nu} \stackrel{\rm def.}{=} {e_i}^\lambda
{e^i}_{\mu, \nu}, \ee where $e_{i \mu, \nu}=\partial_\nu e_{i
\mu}$. The curvature tensor defined by ${\Gamma^\lambda}_{\mu
\nu}$ is identically vanishing, however.  The metric tensor
$g_{\mu \nu}$
 is given by
 \be g_{\mu \nu}= \eta_{i j} {b^i}_\mu {b^j}_\nu, \ee
with the Minkowski metric $\eta_{i j}=\textrm {diag}(+1\; ,-1\;
,-1\; ,-1)$\footnote{ Latin indices are rasing and lowering with
the aid of $\eta_{i j}$ and $\eta^{ i j}$.}.
 We note that, associated with any tetrad field ${e_i}^\mu$ there
 is a metric field defined
 uniquely by (4), while a given metric $g^{\mu \nu}$ does not
 determine the tetrad field completely; for any local Lorentz
 transformation of the tetrads ${e_i}^\mu$ leads to a new set of
 tetrads which also satisfy (4).
  The gravitational Lagrangian ${\cal L_G}$ has the form
\be {\cal L}_G  =  \sqrt{-g} L_G=\sqrt{-g} \left(-\displaystyle{1
\over 3 \kappa}(t^{\mu \nu \lambda}t_{\mu \nu \lambda}-  \Phi^\mu
\Phi_\mu)+\xi a^\mu a_\mu\right),\ee where $t_{\mu \nu \lambda}$,
$ \Phi_\mu$ and $a_\mu$ are  irreducible representation of the
torsion tensor defined by \ba t_{\mu \nu \lambda} \A \stackrel{\rm
def.}{=} \A \displaystyle{1 \over 2} (T_{\mu \nu \lambda}+T_{\nu
\mu \lambda})+\displaystyle{1 \over 6}(g_{\lambda
\mu}\Phi_\nu+g_{\lambda \nu}\Phi_\mu)-\displaystyle{1 \over
3}g_{\mu \nu}\Phi_\lambda,\nonu
\Phi_\mu \A \stackrel{\rm def.}{=} \A {T^\lambda}_{\mu \lambda},
\qquad \qquad a_\mu  \stackrel{\rm def.}{=} \displaystyle{1 \over
6} \epsilon_{\mu \nu \rho \sigma}T_{\nu \rho \sigma}, \ea with
$\epsilon_{\mu \nu \rho \sigma}$ is a totally antisymmetric tensor
normalized to
\[ \epsilon_{0 1 2 3}=-\sqrt{-g}, \qquad \qquad with \qquad \qquad g \stackrel{\rm
def.}{=}  \det.(g_{\mu \nu}),\] and $T_{\mu \nu \lambda}$ is the
torsion tensor defined by
 \be {T^\lambda}_{\mu \nu}\stackrel{\rm
def.}{=}{\Gamma^\lambda}_{\mu \nu}-{\Gamma^\lambda}_{\nu \mu}.\ee
$\kappa$ and $\xi$ are the Einstein gravitational constant and  a
free dimensionless parameter\footnote{Throughout this paper we use
the relativistic units$\;$ , $c=G=1$ and
 $\kappa=8\pi$.}.

 The gravitational field equations for the
system described by ${\it L_G}$ are the following:

 \be G_{\mu \nu}(\{\})
+H_{\mu \nu} = -{\kappa} T_{\mu \nu}, \ee \be
\partial_\mu\left(J^{i j \mu}\right)=0, \ee where the Einstein tensor
$G_{\mu \nu}(\{\})$ is defined by \be G_{\mu
\nu}(\{\})\stackrel{\rm def.}{=} R_{\mu \nu}(\{\})-{1 \over 2}
g_{\mu \nu} R(\{\}), \ee where $R_{\mu \nu}(\{\})$ is the Ricci
tensor and $R(\{\})$ is the Ricci scalar. We
 assume  that the energy-momentum tensor of matter fields is
symmetric. The energy-momentum tensor of a source field with
Lagrangian $L_M$: \be \sqrt{-g} T^{\mu \nu}  \stackrel{\rm
def.}{=}e^{i \mu} \displaystyle {\delta (-\sqrt{-g} L_M) \over
\delta {e^i}_\nu}.\ee Here $H_{\mu \nu}$ and $J_{i j \mu}$ are
given by \be H^{\mu \nu} \stackrel{\rm def.}{=}
\displaystyle{\kappa \over \lambda} \left[\displaystyle{1 \over 2}
\left\{ \epsilon^{\mu \rho \sigma \lambda}({T^\nu}_{\rho
\sigma}-{T_{\rho \sigma}}^\nu)+\epsilon^{\nu \rho \sigma
\lambda}({T^\mu}_{\rho \sigma}-{T_{\rho
\sigma}}^\mu)\right\}a_\lambda-\displaystyle{3 \over 2}a^\mu
a^\nu-\displaystyle{3 \over 4}g^{\mu \nu}a^\lambda a_\lambda
\right], \ee and

\be J^{i j \mu} \stackrel{\rm def.}{=} -\displaystyle{1 \over
2}{e^i}_\rho {e^j}_\sigma \epsilon^{\rho \sigma \mu \nu} a_\nu,\ee
respectively, where \be \lambda \stackrel{\rm def.}{=}
\displaystyle{4 \over 9} \xi+\displaystyle{1 \over 3\kappa}.\ee
Therefore, both $H_{\mu \nu}$ and $J_{i j \mu}$ vanish if the
$a_\mu$ is vanishing. In other words, when the $a_\mu$ is found to
vanish from the antisymmetric part of the field equations, (9),
the symmetric part (8) coincides with the Einstein equation. When
the dimensionless parameter $\lambda=0$ then the theory reduce to
Einstein teleparallel theory equivalent to general relativity.

\newsection{Axially symmetric solution}

In this section we will assume the  parallel vector fields to have
the  form \be \left({b^i}_\mu \right) =\left(\matrix {A(x,y) & 0&
0 & 0 \vspace{3mm} \cr 0 & B(x,y) & 0 & 0 \vspace{3mm} \cr
0&0&C(x,y)&0 \vspace{3mm} \cr 0&0&0&D(x,y) \cr } \right), \ee
where $A(x,y), B(x,y), C(x,y), D(x,y)$ are unknown functions.
Applying (15) to the field equations (8) and (9) one can obtains
the unknown functions in the form \ba \A \A  A(x,y)=\displaystyle{
\sqrt{G(y)} \over A(x+y)}, \qquad \qquad \qquad \qquad
B(x,y)=\displaystyle{1 \over A(x+y)\sqrt{G(x)} }, \nonu
\A \A C(x,y) =\displaystyle{1 \over  A(x+y)\sqrt{G(y)} }, \qquad
\qquad \qquad D(x,y)=\displaystyle{\sqrt{G(x)} \over A(x+y)}, \ea
where $G(\xi)=1-{\xi}^2-2mA{\xi}^3,$ $A$ and $m$ are positive
parameters related to the mass and acceleration of the black hole
and satisfying $Am<1/\sqrt{27}$ \cite{KE}. The associated metric
of solution (16) has the form (1) which is the C-metric. As it is
stated in the introduction that in general one can not easily
write down simple expression of the roots of $G$. Therefore, one
must find some coordinate transformation which makes the roots  of
$G$ written explicitly and this would in turn simplify certain
analysis of the C-metric. This coordinate transformation has the
form \cite{KE} \ba x \A=\A \displaystyle{ \sqrt{1+6{\bar m}{\bar
A} c_1 \over 1+{c_1}^2+4{\bar m} {\bar A} {c_1}^3} }\left({\bar
x}-c_1\right), \qquad y = \displaystyle{ \sqrt{1+6{\bar m}{\bar A}
c_1 \over 1+{c_1}^2+4{\bar m} {\bar A} {c_1}^3} }\left({\bar
y}-c_1\right), \nonu
\phi \A=\A  \sqrt{(1+6{\bar m}{\bar A} c_1) (1+{c_1}^2+4{\bar m}
{\bar A} {c_1}^3) }{\bar \phi}, \quad t = \sqrt{(1+6{\bar m}{\bar
A} c_1) (1+{c_1}^2+4{\bar m} {\bar A} {c_1}^3) }{\bar t},\ea where
${\bar x}, {\bar y}, {\bar \phi}, {\bar t}$ are the new coordinate
and \be m={{\bar m} \over (1+6{\bar m}{\bar A} c_1)^{3/2}}, \qquad
\qquad A=\sqrt{(1+{c_1}^2+4{\bar m}{\bar A} {c_1}^3)}{\bar A}. \ee
Applying the coordinate transformation (17) to the tetrad (15)
with solution (16) we obtain \be  \left({b^i}_\mu \right)
=\left(\matrix { \displaystyle{G({\bar y}) \over {\bar A} ({\bar
x} -{\bar y})} & 0 & 0  & 0 \vspace{3mm} \cr 0 & \displaystyle{1
\over {\bar A} ({\bar x} -{\bar y}) G({\bar x})}  & 0& 0
\vspace{3mm} \cr 0&0&\displaystyle{1 \over {\bar A} ({\bar x}
-{\bar y})G({\bar y})} &0 \vspace{3mm} \cr 0&0&0& \displaystyle{
G({\bar x}) \over {\bar A} ({\bar x} -{\bar y})} \cr } \right),
\ee  with the structure function defined by \be G(\xi)
\stackrel{\rm def.}{=} (1-\xi^2)(1+2{\bar m} {\bar A} \xi)\ee  and
the associated metric has the form (1) with the structure function
has the form (20). As is clear from (20) that one can gets the
roots easily which has the form \be \xi_{1,2}=\pm 1, \quad
\xi_3=-\displaystyle{ 1 \over 2{\bar m} {\bar A}}, \quad which
\quad obey \quad \xi_3<\xi_2<\xi_1.\ee Now we are going to study
the physics of tetrad (19) by studying the singularities.
\newsection{Singularities}

In teleparallel theories we mean by singularity of spacetime
\cite{Kn} the singularity of the scalar concomitants  of the
curvature and torsion tensors.

Using the definitions  of the  Riemann Christoffel, Ricci tensors,
Ricci scalar, torsion tensor, basic vector, traceless part and the
axial vector part \cite{Ncs} we obtain for the  solution (19)
 \ba
 R^{\mu \nu \lambda \sigma}R_{\mu \nu \lambda
\sigma} \A = \A 48{\bar A}^6({\bar x}-{\bar y})^6{\bar m}^2,
\qquad R^{\mu \nu}R_{\mu \nu}=0, \qquad  R=0,\nonu
T^{\mu \nu \lambda}T_{\mu \nu \lambda} \A=\A \displaystyle{
F_1({\bar x},{\bar y}) \over (1-{\bar x}^2)(1+2{\bar x}{\bar
m}{\bar A})(1-{\bar y}^2)(1+2{\bar y}{\bar m}{\bar A})}, \nonu
\Phi^\mu \Phi_\mu \A=\A  \displaystyle{F_2({\bar x},{\bar y})
\over (1-{\bar x}^2)(1+2{\bar x}{\bar m}{\bar A})(1-{\bar
y}^2)(1+2{\bar y}{\bar m}{\bar A})} ,\nonu
t^{\mu \nu \lambda}t_{\mu \nu \lambda} \A=\A
\displaystyle{F_2({\bar x},{\bar y}) \over (1-{\bar x}^2)(1+2{\bar
x}{\bar m}{\bar A})(1-{\bar y}^2)(1+2{\bar y}{\bar m}{\bar A})}
,\nonu
a^\mu a_\mu\A =\A 0.
 \ea
 As  is clear from (22) that the scalars of torsion, basic vector
 and traceless part have the same singularities, let us discuss
 these singularities.\\ 1) When ${\bar x}={\bar y}=\xi_2$ then all the scalars of (22)
 have a singularities which is called {\it asymptotic infinity} \cite{KE}
  .\\ 2) When ${\bar y}=\xi_3$, there is a singularity
 which is called {\it black hole event horizon} \cite{KE}.\\3)
 When ${\bar y}=\xi_2$ there is also a singularity which is {\it acceleration
 horizon}.\\ 4) When ${\bar x}=\xi_1$ there is a singularity which makes
 {\it symmetry axis between event and acceleration horizons}.\\ 5) When
 ${\bar x}=\xi_2$ there is a singularity which makes {\it a symmetry axis
 joining between event horizon with asymptotic horizon}.\\ 6) When
 ${\bar x}=\xi_2$ and ${\bar y}=\xi_1$ there will be  {\it a conical singularity} \cite{KE}.
\newsection{Energy content }
Now we are going to use the following coordinate transformation
\cite{KE} \be {\bar x}=\cos(\theta), \qquad \qquad {\bar
y}=-\displaystyle{1 \over {\bar A} r}, \qquad {\bar t}={\bar
A}t_1, \ee where $r, \theta, t_1$ are the new coordinate. Applying
transformation (23) to tetrad (19) we get \be \left({b^i}_\mu
\right) =\left(\matrix {-{ \sqrt{r-2{\bar m}}\sqrt{{\bar
A}^2r^2-1} \over \sqrt{r}({\bar A}r \cos \theta+1)}&0 &0 &0
\vspace{3mm} \cr 0&0 &- { r \over \sqrt{1+2{\bar m} {\bar A}\cos
\theta}({\bar A}r \cos \theta +1)} &0 \vspace{3mm} \cr 0& -{
\sqrt{r} \over \sqrt{r-2{\bar m}}\sqrt{{\bar A}^2r^2-1}({\bar A}r
\cos \theta+1)}&0&0 \vspace{3mm} \cr 0&0&0& {\sin \theta
\sqrt{1+2{\bar m} {\bar A}\cos \theta} \over {\bar A}r \cos
\theta+1} \cr } \right). \ee Taking the $limit_{{\bar A}
\rightarrow 0}$ in (24), the associate metric will have the
Schwarzschild form. Now we are going to study some physics of this
solution by calculate its energy content.

 The superpotential is given by \be {{\cal U}_\mu}^{\nu \lambda} ={(-g)^{1/2} \over
2 \kappa} {P_{\chi \rho \sigma}}^{\tau \nu \lambda}
\left[\Phi^\rho g^{\sigma \chi} g_{\mu \tau}
 -\lambda g_{\tau \mu} \gamma^{\chi \rho \sigma}
-(1-2 \lambda) g_{\tau \mu} \gamma^{\sigma \rho \chi}\right], \ee
where ${P_{\chi \rho \sigma}}^{\tau \nu \lambda}$ is \be {P_{\chi
\rho \sigma}}^{\tau \nu \lambda} \stackrel{\rm def.}{=}
{{\delta}_\chi}^\tau {g_{\rho \sigma}}^{\nu \lambda}+
{{\delta}_\rho}^\tau {g_{\sigma \chi}}^{\nu \lambda}-
{{\delta}_\sigma}^\tau {g_{\chi \rho}}^{\nu \lambda} \ee with
${g_{\rho \sigma}}^{\nu \lambda}$ being a tensor defined by \be
{g_{\rho \sigma}}^{\nu \lambda} \stackrel{\rm def.}{=}
{\delta_\rho}^\nu {\delta_\sigma}^\lambda- {\delta_\sigma}^\nu
{\delta_\rho}^\lambda. \ee The energy is expressed by the surface
integral \cite{Mo2} \be E=\lim_{r \rightarrow
\infty}\int_{r=constant} {{\cal U}_0}^{0 \alpha} n_\alpha dS, \ee
where $n_\alpha$ is the unit 3-vector normal to the surface
element ${\it dS}$.

Now we are in a position to calculate the energy associated with
solution (24) using the superpotential (25). As is
 clear from (28), the only components which contributes to the energy is ${{\cal U}_0}^{0
 \alpha}$. Thus substituting from solution (24) into
 (25) we obtain the following non-vanishing value
 \be
{{\cal U}_0}^{0 a}=-{x^a \over \kappa
r^3(x^2+y^2)}\left(r^3+(r-4M)(x^2+y^2) \right), \qquad a=1,2,
\qquad {{\cal U}_0}^{0 3}=-{z \over \kappa r^3}\left(r-4M \right).
 \ee
 Substituting from (29) into
(28) we get \be E(r)=2M-r! \ee
\newsection{Main results and Discussion}

 We begin with tetrad (15) which is the square root of metric (1).
 We then, preform the coordinate transformation (17) to tetrad
 (15) with solution (16). The associated metric of tetrad (19)
 has the structure function (20), that is offers advantage over the
 traditional form (1). With a factorisable structure function, its
 roots can be explicitly read off from it, leading to simplifications
  in the analysis of the C-metric. Besides the simplifying known results,
the new form of the C-metric opens up the possibility of
performing calculations that were previously impractical if not
possible. We also study the singularities of tetrad (19) as we see
from (22) that we easily find the singularities which are the
roots of the structure function (20).

Perform the coordinate transformation (23) to the tetrad (19), we
obtain the tetrad (24) whose associated metric is the
Schwarzschild spacetime. Calculating the energy content of this
solution we obtain (30) which is a very strange results! In fact
if we take the limit of (30) when $r\rightarrow \infty$ we will
get a meaningless result. We can offer the following reasons for
such strange result\vspace{.3cm}\\i) Is the tetrad (15) that we
begin with it its structure is inconsistence? If this is true this
means that the square root of metric that will constitute the
tetrad is not a physical one.\vspace{.3cm}\\ ii) Another
possibility for this strange result is that the transformation
(17) and (23) makes the tetrad (15) in its final form (24)
unphysics!

 In the present time it is important to find a consistence tetrad
 which can reproduce the structure function (1), then preforming
 the coordinate transformation (17) and (23). Calculations of
  the energy content of the tetrad that reproduce the
 Schwarzschild spacetime  must be coincide with inertial
 mass. Work in this direction is in progress.

\end{document}